\documentclass[aps,showpacs]{revtex4}
\begin{document}
\title{Weyl electrovacuum solutions and gauge invariance}
\author{Boyko V. Ivanov}
\email{boyko@inrne.bas.bg}
\affiliation{Institute for Nuclear Research and Nuclear Energy,\\
Tzarigradsko Shausse 72, Sofia 1784, Bulgaria}

\begin{abstract}
It is argued that in Weyl electrovacuum solutions the linear term in the
metric cannot be eliminated just on grounds of gauge invariance. Its
importance is stressed.
\end{abstract}

\maketitle

In general relativity static electric fields alter the metric of spacetime
through their energy-momentum tensor \cite{one} 
\begin{equation}
T_{\;\nu }^\mu =\frac \varepsilon {4\pi }\left( F^{\mu \alpha }F_{\nu \alpha
}-\frac 14\delta _{\;\nu }^\mu F^{\alpha \beta }F_{\alpha \beta }\right) ,
\label{one}
\end{equation}
where 
\begin{equation}
F_{\mu \nu }=\partial _\mu A_\nu -\partial _\nu A_\mu  \label{two}
\end{equation}
is the electromagnetic tensor, $A_\mu =\left( \bar \phi ,0,0,0\right) $ is
the four-potential and $\varepsilon $ is the dielectric constant of the
medium. $T_{\;\nu }^\mu $ enters the r.h.s. of the Einstein equations 
\begin{equation}
R_{\;\nu }^\mu =\kappa T_{\;\nu }^\mu ,  \label{three}
\end{equation}
where $\kappa $ is the Einstein constant. We have taken into account that $%
T_{\;\mu }^\mu =0$. In addition, the Maxwell equations are coupled to
gravity through the covariant derivatives of $F_{\mu \nu }$%
\begin{equation}
F_{\quad ;\nu }^{\mu \nu }=\frac 1{\sqrt{-g}}\left( \sqrt{-g}F^{\mu \nu
}\right) _\nu =0,  \label{four}
\end{equation}
where $g$ is the metric's determinant and usual derivatives are denoted by
subscripts. The electric field is $E_\mu =F_{0\mu }=-\bar \phi _\mu $ .
Obviously, $T_{\;\nu }^\mu $ from Eq. (1) contains only quadratic terms in $%
\bar \phi _\mu $. This allows to hide $\kappa $ and $\varepsilon $ by
normalizing the electric potential to a dimensionless quantity 
\begin{equation}
\phi =\sqrt{\frac{\kappa \varepsilon }{8\pi }}\bar \phi .  \label{five}
\end{equation}
The factor $8\pi $ is chosen for future convenience and we use CGS units.
This is a much more efficient way to get rid of the constants in the
Einstein-Maxwell equations than the choice of relativistic units.

Let us confine ourselves to the axially-symmetric static metric \cite{two} 
\begin{equation}
ds^2=f\left( dx^0\right) ^2-f^{-1}\left[ e^{2k}\left( dr^2+dz^2\right)
+r^2d\varphi ^2\right] ,  \label{six}
\end{equation}
where $x^0=ct$, $x^1=\varphi ,$ $x^2=r,$ $x^3=z$ are cylindrical
coordinates, $f=e^{2u}$ and $u$ is the first, while $k$ is the second
gravitational potential. Both of them depend only on $r$ and $z$. For the
electric field one has 
\begin{equation}
E_r=-\bar \phi _r,\qquad E_z=-\bar \phi _z.  \label{seven}
\end{equation}
The field equations read 
\begin{equation}
\Delta u=e^{-2u}\left( \phi _r^2+\phi _z^2\right) ,  \label{eight}
\end{equation}
\begin{equation}
\Delta \phi =2\left( u_r\phi _r+u_z\phi _z\right) ,  \label{nine}
\end{equation}
\begin{equation}
\frac{k_r}r=u_r^2-u_z^2-e^{-2u}\left( \phi _r^2-\phi _z^2\right) ,
\label{ten}
\end{equation}
\begin{equation}
\frac{k_z}r=2u_ru_z-2e^{-2u}\phi _r\phi _z,  \label{eleven}
\end{equation}
where $\Delta =\partial _{rr}+\partial _{zz}+\partial _r/r$ is the
Laplacian. We have used the definition given in Eq. (5). The first two
equations determine $\phi $ and $f$. After that $k$ is determined by
integration.

Weyl electrovacuum solutions \cite{three} are obtained when the
gravitational and the electric potential have the same equipotential
surfaces, $f=f\left( \phi \right) $. Eqs. (8-9) yield 
\begin{equation}
\left( f_{\phi \phi }-2\right) \left( \phi _r^2+\phi _z^2\right) =0,
\label{twelve}
\end{equation}
which gives 
\begin{equation}
f=A+B\phi +\phi ^2,  \label{thirteen}
\end{equation}
where $A$ and $B$ are arbitrary constants. Replacing it in Eqs. (8-9) one
comes to an equation for $\phi $%
\begin{equation}
\Delta \phi =\frac{B+2\phi }{A+B\phi +\phi ^2}\left( \phi _r^2+\phi
_z^2\right) .  \label{fourteen}
\end{equation}
Let us make one more assumption, that $\phi $ depends on $r,z$ through some
function $\psi \left( r,z\right) $ which satisfies the Laplace equation $%
\Delta \psi =0$. Then $\phi \left( \psi \right) $ is determined implicitly
from 
\begin{equation}
\psi =\int \frac{d\phi }{A+B\phi +\phi ^2}.  \label{fifteen}
\end{equation}
An important equality follows 
\begin{equation}
\phi _i=f\psi _i,\qquad \bar \phi _i=f\left( \phi \right) \bar \psi _i,
\label{sixteen}
\end{equation}
where $i=r,z$. Eqs. (10-11) become 
\begin{equation}
k_r=\frac D4r\left( \psi _r^2-\psi _z^2\right) ,\qquad k_z=\frac D2r\psi
_r\psi _z,  \label{seventeen}
\end{equation}
where $D=B^2-4A$. Thus in Weyl electrovac solutions the harmonic master
potential $\psi $ determines the electric and the gravitational fields.

The theory should be invariant under gauge transformations, which in this
case are simply translations: $\phi ^{\prime }=\phi +a$ with $a$ being an
arbitrary constant. Eqs.(1-4) and (7-11) are gauge invariant, but Eq.(13) is
not because $A,B$ change into 
\begin{equation}
A^{\prime }=A+Ba+a^2,\qquad B^{\prime }=B+2a.  \label{eighteen}
\end{equation}
This happens because $f$ depends directly on the electric potential and not
on its derivatives. In some papers this is used to set $B^{\prime }$ to zero
and eliminate the linear term.

In this paper we shall show that this is not correct. In fact, the general
solution (13) stays gauge invariant because $A^{\prime },B^{\prime }$ are
also arbitrary constants. In a particular solution $A^{\prime },B^{\prime }$
should be fixed and should not change under a gauge transformation. This is
possible when after $a$ is selected one compensates its effect by choosing $%
A,B$ in such a way that $A^{\prime },B^{\prime }$ stay fixed at any
particular value. Eq.(18) shows that this always can be done and in this way
the gauge invariance of $f$ is restored. For example, due to Eq.(5), the
electric potential is very small everywhere for realistic fields and it is
natural that it should go to zero at infinity or when the field is turned
off. Then asymptotic flatness requires to set $A^{\prime }=1$ and this
condition can be kept in spite of possible gauge transformations. The
coefficient $B^{\prime }$ is not determined by the system of equations (8-9)
and the Weyl conditions. One can not just put it to zero by a gauge
transformation. In fact, arguments were given in \cite{four,five} that its
value is $2$. Then $f$ becomes a perfect square, while $k$ vanishes and the
space part of the metric is conformally flat. It should be noticed that $%
D^{\prime }=D$ so that the vanishing of $k$ is gauge invariant.

The presence of the linear term in $f$ with a coefficient of order unity is
not just of academic interest. Because of the gravitational potential a
particle at rest feels an acceleration \cite{one} 
\begin{equation}
g_i=\frac{c^2}2\left( \ln g_{00}\right) _i=c^2f^{-1}\left( \frac{B^{\prime }}%
2\sqrt{\frac{\kappa \varepsilon }{8\pi }}\bar \phi _i+\frac{\kappa
\varepsilon }{8\pi }\bar \phi \bar \phi _i\right) .  \label{nineteen}
\end{equation}
Covariant and contravariant components coincide in practice because for
realistic electric fields the metric is almost flat. Eq.(7) shows that the
first term is proportional to the electric field, which due to Eq.(16) may
be derived also from the master potential because $f$ is extremely close to
one. Let us note that

\begin{equation}
c^2\sqrt{\frac \kappa {8\pi }}=\sqrt{G}=2.58\times 10^{-4},\qquad c^2\frac 
\kappa {8\pi }=\frac G{c^2}=7.37\times 10^{-27},  \label{twenty}
\end{equation}
where $G=6.674\times 10^{-8}cm^3/g.s^2$ is the Newton constant and $%
c=2.998\times 10^{10}cm/s$ is the speed of light. Due to the square root,
the first coefficient is $10^{23}$ times bigger than the second and for
realistic fields and media this cannot be compensated by the squares of
potentials and the additional $\sqrt{\varepsilon }$ factor in the second
term. The latter is typical for linear perturbation theory. In relativistic
units $G=c=1$ the difference does not show up. Thus, provided that $%
B^{\prime }=2$, the linear term is essential and the coupling of
electromagnetism to gravity appears to be much stronger than it is usually
thought. It causes a number of effects, the most prominent being the
movement of a usual capacitor towards one of its poles. In this case there
is plane symmetry in the bulk, $f$ and $\phi $ depend only on $z$, which
means they are functionally related and the general solution belongs to the
Weyl class. However, Eqs.(10,17) show that $k$ depends on $r$ and breaks the
symmetry unless $D=0$, which gives $B^{\prime }=\pm 2$. Putting the usual
formula for the electric field inside a capacitor into Eq.(19) gives for the
acceleration which acts on the dielectric inside it 
\begin{equation}
g_z=\pm \sqrt{G\varepsilon f}\frac{\bar \psi _0}d\approx \pm 2.58\times
10^{-4}\frac{\sqrt{\varepsilon }}d\bar \psi _0,  \label{twentyone}
\end{equation}
where $\psi _0$ is the potential difference between the plates and $d$ is
the distance between them. A more detailed derivation can be found in Refs.%
\cite{four,five}. If the capacitor is hanging freely, this effect may be
tested experimentally. To increase the acceleration it is advantageous to
make $d$ small (typically $0.1cm\leq d\leq 1cm$), to raise $\psi _0$ up to $%
2\times 10^4CGS$ (six million volts, which is possible) and to take a
ferroelectric material with $\varepsilon $ in the range of $10^4$, like
barium titanate ($BaTiO_3$) or many others. Thus $\sqrt{\varepsilon }/d$ may
reach in principle $10^3$ and the maximum acceleration $g_{z,\max
}=5.2g_{earth}$ is more than enough to counter Earth's gravity.

This effect has been discovered by the prominent electrical engineer Thomas
Townsend Brown (1905-1985) already in 1923 together with Prof. P. A. Biefeld
and called the Biefeld-Brown effect \cite{six}. Brown worked on his own on
it up to the sixties with high voltage equipment in the range $70-300kV.$ He
didn't give a formula like Eq.(21) but stressed that the effect is bigger
the closer the condenser plates, the higher the voltage and the greater the $%
\varepsilon ,$ which is in accord with Eq.(21). He also found that the
capacitor moves towards its positive pole, resolving experimentally the sign
ambiguity in the above formula. There have been speculations that the effect
might follow from some of the Einstein's unified theories. Today one would
mention string theory or some other alternative gravitational theory.
However, it appears that the effect is a part of usual General Relativity
due to its strong nonlinearity. It is worth to repeat Brown's experiments in
different laboratories and check formula (21).

\end{document}